\documentclass[11pt]{article}

\usepackage{graphicx}

\usepackage{doublespace}

\usepackage{amssymb}
\usepackage{amsmath}
\usepackage{amsthm}

\usepackage{xspace,colortbl}

\def\ep{\epsilon}

\def\ep{\epsilon}
\def\a{\alpha}

\def\th{\theta}
\def\*{\cdot}

\def\d{\partial}

\def\ra{\rightarrow}
\def\vp{\varphi}

\def\La{\Lambda}
\def\lla{\frac{\Lambda}{3}}
\def\ka{\kappa}

\def\Q{\mathcal Q}
\def\P{\mathcal P}

\newcommand\s{\sigma}

\definecolor{dark_blue}{rgb}{0,0,0.7}

\begin{document}

\title{A vortex model for rotating compact objects}

\author{G. F. Chapline\footnote{Lawrence Livermore National Laboratory, Livermore, CA 94550}, P. Marecki\footnote{Institut f\"ur Theoretische Physik, D-04009 Leipzig, Germany}}

 \maketitle

\begin{abstract}
In this paper a rotating stationary solution of the vacuum Einstein equations with a cosmological constant is exhibited which reduces to de Sitter's interior cosmological solution when the angular momentum goes to zero. This solution is locally isomorphic to de Sitter space, but as one approaches the axis of rotation it has a novel feature: a conical event horizon. This suggests that in reality rotating compact objects have a vortex structure similar to that conjectured for rotating superfluid droplets. In the limit of slow rotation the vortex core would be nearly cylindrical and the  space-time inside the core would be G\"odel-like. The exterior space-time will resemble the Kerr solution for equatorial latitudes, but significant deviations from Kerr are expected for polar latitudes.
\end{abstract}

\section{Introduction}
$\qquad$ In this note we exhibit an exact solution to the Einstein field equations that may shed light on two outstanding problems in theoretical astrophysics. The first problem is to describe the {\it physical nature} of rotating compact objects (i.e. objects that within the framework of general relativity would lie within an event horizon). This is very deep problem because the classical interior Kerr solution has a number of unphysical features \cite{SourceKerr}, and so finding a model for the interior space-time of a compact object necessarily involves violating general relativity. In the case of a non-rotating compact object it has been proposed \cite{CHHLS,MM} that the interior space-time can be modeled using de Sitter's 1917 ``interior'' static cosmological solution \cite{deSitter}. This provides an additional enigma though because it has been an open question as to how to generalize de Sitter's solution to include rotation. Our investigation of the nature of rotating compact objects begins by providing a direct answer to this latter question. Our rotating version of de Sitter's interior static solution in turn suggests a quite attractive picture for the physical structure of the interior space-time of a rotating compact object.

The problem of describing any rotating object (even a rotating planet!) within the framework of general relativity also runs into the difficulty that there is no universal time for such solutions (in other words the classical space-times are asynchronous). Indeed, this was one of the main motivations for G\"odel in introducing his famous model for a rotating universe \cite{Goe}. This means that such space-times are inconsistent with elementary quantum mechanics, because quantum mechanics - at least in the form described in standard textbooks - requires a universal time for its definition. Another problem with the lack of a universal time for rotating solutions to the Einstein equations is that closed time-like curves often appear, which creates problems with causality.

Recently it has been suggested \cite{CHM04} that the way to resolve the difficulties with rotating solutions of the classical Einstein equations is to suppose that the rotation is actually carried by spinning space-time strings \cite{Mazur}, in a manner analogous to the way rotation of superfluid helium inside a rotating container is carried by quantized vortices. The spinning strings resolve the question of the consistency of rotating space-times with quantum mechanics because the vorticity of space-time would be concentrated into the cores of the spinning strings (where general relativity fails). Our proposed model for a rotating compact object is a generalization of the space-time spinning string in flat space-time to the case of a finite rotating object. As in the case of the spinning string the axis of rotation sits inside a ``hole'' in the surrounding space-time.

In the picture of non-rotating compact objects suggested in Ref's \cite{CHHLS,MM} the outer edge of de Sitter's interior static solution edge is matched to the external Schwarzschild solution via a surface layer where general relativity fails. One nagging question concerning this proposal, though, has been the question as to what should replace de Sitter's interior solution in the case of a rotating compact object.  Although our rotating de Sitter solution is not mathematically matched to the exterior Kerr solution at the Kerr event horizon, we will present a number of arguments to the effect that it provides the sought extension of the Kerr spacetime to (at least some regions of) the interior of the compact object, and in turn suggests an attractive model for the structure of a compact rotating object. Where we are led can be explained with the help of the condensate model for non-rotating compact objects discussed in \cite{CHHLS}: while a collapsed non-rotating object consists of a spherical condensate surrounded by the usual vacuum, a rotating compact object will consist of an ``apple-shaped'' condensate containing a vortex-like core surrounding its axis of rotation. As in the non-rotating case we will assume that the condensate corresponds to a region with a large vacuum energy, but no ordinary matter. Curiously our model for rotating compact objects resembles Zeldovich's model for a rotating superfluid droplet \cite{Zeldovich}.

\section{The metric}
$\qquad$ The ``rotating de Sitter'' metric which we propose as a model for the interior of a rotating compact object is (we use units such that $8\pi G/c^{2} = 1$)
\begin{align}
\nonumber
ds^2&=\left[1-\lla(r^2-a^2\cos^2\th)\right]\, dt^2+2a\lla \left[r^2-(r^2+a^2)\cos^2\th\right]\, dt\, d\vp - \frac{\rho^2}{r^2+a^2-\lla r^4} \, dr^2\\
&-\frac{\rho^2}{1-\lla a^2 \cos^2\th\cot^2\th} \, d\th^2
-\left\{ (r^2+a^2)\sin^2\th+\lla a^2 [r^2-(a^2+2r^2)\cos^2\th]  \right\}\, d\vp^2\label{metric}
\end{align}
where $a$ is the angular momentum per unit mass,  $\rho^2=r^2+a^2\cos^2\th$. This metric is a limiting case of a rich class of metrics discovered by Carter \cite{Carter} and independently by Plebanski \cite{Plebanski}. While a number of simpler solutions can be derived from the Carter-Plebanski metric by various limiting procedures (see e.g. \cite{GriffPod}), none of these has so far (to our knowledge) found application to the astrophysics of compact objects. In the limit $a\ra 0$ our metric \eqref{metric} reduces to de Sitter's 1917 metric. The apparent singularities in the $g_{rr}$ and $g_{\th\th}$ components of the metric tensor can be removed by a change of variables \cite{Plebanski}, and represent surfaces where $g_{00}g_{\vp\vp}- g_{0\vp}^2 = 0$. The singularity in $g_{rr}$ is associated with a spherical surface at
\begin{equation}\label{spherical_horizon}
 r_H^2=\frac{3}{2\La}\left(1+\sqrt{1+\lla a^2} \right)
\end{equation}
In the limit $a\ra 0$ this surface becomes the edge of the interior de Sitter space-time for a non-rotating object. In addition to the spherical event horizon \eqref{spherical_horizon} there is a conical event horizon located at
\begin{equation}
 \label{conical_horizon}
\tan^2\th_H=-\frac{1}{2}+\sqrt{\lla a^2+\frac{1}{4}}
\end{equation}
In the case of slow rotation $\La a^2\ll 1$ the conical event horizon is located at very high latitudes.

Closed time-like curves, which as mentioned above is a common disease of rotating solutions of the Einstein eq's., will appear if $g_{\vp\vp}$ is positive for some values of $r$. In our case this means that
\begin{equation}\label{CTCs}
 (r^2+a^2)\sin^2\th+\lla a^2 [r^2-(a^2+2r^2)\cos^2\th] <0.
\end{equation}
This condition will be satisfied only for very high latitudes (for slow rotation). This is very reminiscent of the situation with spinning strings \cite{Mazur}.
The relative position of the region with CTCs and the conical horizon \eqref{conical_horizon} is shown in Fig 1. Note, that the meaning of  ``the radius'' (the coordinate $r$) here is provided by the circumference of circles $r=const$, $\th=const$.

Let us now attempt to compare the solution \eqref{metric} to the Kerr solution, which presumably provides an approximately correct description of space-time outside the of the event horizon \eqref{spherical_horizon}. The circumference of an equatorial circle at the surface \eqref{spherical_horizon} will be equal to the circumference of the equatorial circle on the outer horizon of Kerr space-time if the mass parameter for the Kerr solution fulfills
\begin{equation}
 2m=\lla r_H^3.
\end{equation}
Curiously this is the same condition that was used in Ref's \cite{CHHLS,MM} to match de Sitter's interior solution to the exterior Schwarzschild solution in the case of a non-rotating compact object. We note that the value $\lla a^2=2$ is critical, as $m\geqslant a$ is true for all values of $\lla$ with quality occurring only at $\lla a^2=2$. Near the surface \eqref{spherical_horizon} our rotating metric \eqref{metric} has the form
\begin{equation}
 ds^2=-a^2\, \frac{\sin^2\th -\lla a^2 \cos^4\th}{\rho_H^2}\left(dt-\frac{r_H^2+a^2}{a}\, d\vp\right)^2-\frac{\rho_H^2}{1-\lla a^2\cos^2\th\cot^2\th}\, d\th^2,
\end{equation}
with $\rho_H^2=r_H^2+a^2\cos^2\th$. For comparison the angular part of the Kerr metric when expressed in Boyer-Lindquist coordinates \cite{BL} and evaluated on the event horizon is
\begin{equation}
ds^2=-a^2\, \frac{\sin^2\th}{\rho_H^2}\left(dt-\frac{r_H^2+a^2}{a}\, d\vp\right)^2-\rho_H^2\, d\th^2.
\end{equation}
It can be seen that except near to $\th= 0$ the angular part of our metric near to the spherical event horizon is not too different to the angular part of the Kerr metric at the event horizon for all values of $a$ such that $\La a^2\ll 1$. Significant differences do appear near to $\th = 0$, which we interpret as a non-Kerr like behavior at high latitudes.

Inside the spherical event horizon \eqref{spherical_horizon} the behavior of our metric is completely different from that of the interior Kerr metric. In the case of the Kerr solution $g_{00}<0$ every\-where inside the ``ergosphere'' whose outer boundary lies outside the event horizon at \mbox{$r^2+a^2\cos^2\th=2mr$}. Although our $g_{00}$ is negative at the event horizon (and close to the Kerr $g_{00}$), it is actually positive for all values of $r$ inside \mbox{$r^2=1/\lla+a^2\cos^2\th$}, which for small $\La a^2$ would be almost everywhere in the interior. In addition, in contrast with the Kerr solution the radial metric coefficient $g_{rr}$ is negative for all values of $r$ inside the spherical event horizon. Thus the problematic reversal of the roles of time and radial distance in the interior Kerr solution is avoided.

The Kerr solution has the property that inside the ergosphere static $\d_t^a$-observers cannot exist and all stationary observers necessarily rotate about the axis. At the event horizon the zero-angular-momentum frame rotates with the ``frame dragging'' angular velocity
\begin{equation}
 \left.\frac{d\vp}{dt}\right|_{r=r_H}=\frac{a}{r_H^2+a^2}.
\end{equation}
For the metric \eqref{metric} $g_{00}<0$ at the spherical event horizon and the frame rotation velocity is also $a/(r_H^2+a^2)$, so particles in our space-time must also rotate as they approach the event horizon from the inside. Indeed our interior metric contains a reflection of the usual exterior Kerr ergosphere, with an inside boundary at $r^2=1/\lla +a^2\cos^2\th$. Thus in our picture of rotating compact objects the metric just inside the event horizon is a reflection of the metric just outside, at least away from $\th= 0$. Reflection symmetry between the inner and outer metrics at an event horizon is just the matching condition for metrics suggested earlier for non-rotating compact objects \cite{CHHLS}, and is a consequence of replacing the smooth geometry at an event horizon that is predicted by classical general relativity with a quantum critical layer.

\begin{center}
  \includegraphics[scale=0.85]{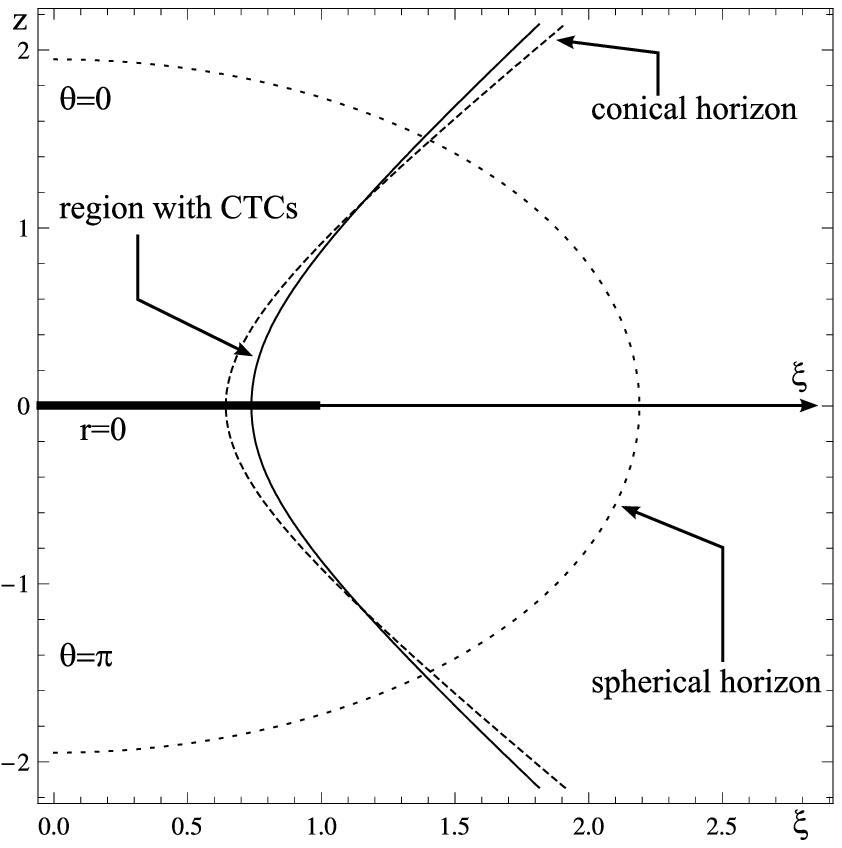}
\end{center}
{\it Fig. 1. Toroidal geometry of the condensate phase of the interior of a rotating compact object. The figure represents a map of constant $\vp$ and $t$ space with pseudocylindrical coordinates $(z,\xi)$,  \mbox{$\xi=\sqrt{r^2+a^2}\, \sin\th $} and $z=r\cos\th$ as in Fig. 1. of \cite{SourceKerr}. The boundary of the region with CTCs is shown in solid, while the conical/spherical horizons are shown dashed/dotted respectively and provide approximate interior/exterior boundaries of the condensate. We anticipate the region containing closed time-like curves (CTCs) to be replaced by a G\"odel-like space-time, while the exterior of the object should be approximately described by the Kerr vacuum solution of Einstein's equations (at least away from the poles).}

\newpage
\section{Comparison with Demianski and Plebanski metrics}
$\qquad$ One might guess that the metric \eqref{metric} could also be derived from Demianski's well known generalization of the Kerr solution to include a cosmological constant \cite{Demianski}. Indeed taking the $m=0$ limit of Demianski's metric\footnote{Further rescalings of $t$ and $\vp$ are necessary for the Demianski spacetime to have a regular axis of rotation, see e.g. \cite{GriffPod}.} yields
\begin{align}\nonumber
 ds^2&=\left[1-\lla (r^2+a^2\sin^2\th)\right]\, dt^2
+2a\lla \sin^2\th (r^2+a^2) \, dt\, d\vp -
\frac{\rho^2}{(r^2+a^2)(1-\lla r^2)}\, dr^2-\\
&-\frac{\rho^2}{1+\ka \cos^2\th}\, d\th^2 -(r^2+a^2)(1+\ka)\sin^2\th d\vp^2,\label{Demianski}
\end{align}
where the parameter $\ka=\lla a^2$ is dimensionless. As was the case for metric \eqref{metric} the variables $\th$ and $\vp$ represent the polar angles on a sphere. At first sight \eqref{metric} and \eqref{Demianski} appear to be different. However, both metrics can be obtained as special cases of the Plebanski metric \cite{Plebanski}, which has the general form
\begin{equation}
 ds^2=\frac{\Q}{p^2+q^2}(d\tau-p^2 d\s)^2-\frac{\P}{p^2+q^2}(d\tau+q^2d\s)^2-\frac{p^2+q^2}{\P}dp^2 -\frac{p^2+q^2}{\Q} dq^2.\label{Plebanski}
\end{equation}
When the mass, NUT charge, and electric and magnetic charges are all zero then the functions $\P(p,q)$ and $\Q(p,q)$ have the simple forms
\begin{equation}
 \P=b-\ep p^2-\lla p^4, \qquad \Q=b+\ep q^2 -\lla q^4.
\end{equation}
For both metrics \eqref{metric} and \eqref{Demianski}
\begin{equation}
 p=a\cos\th, \quad q=r,\quad \s=-\vp/a,\quad \tau=t+a^2\s, \quad b=a^2.
\end{equation}
However, for metric \eqref{metric} $\ep=1$, while for metric \eqref{Demianski} $\ep=1-\ka$.

Evidently the essential difference between the two geometries lies in the value of $\ep$. However it is known that Plebanski metrics with different values of  $\ep$ are related by a certain scaling transformation. This scaling transformation has the form
\begin{equation}
 p'=\frac{p}{\a},\quad q'=\frac{q}{\a},\quad \s'=\s\cdot \a^3,\quad \tau'=\tau\cdot \a,\quad b'=\frac{b}{\a^4}, \quad \ep'=\frac{\ep}{\a^2},\label{scaling}
\end{equation}
where $\a$ is the scaling parameter. The value of $\lla$ is unchanged.  Let the parameters corresponding to the Demianski  metric from now on be distinguished by the subscript $D$. That the metrics \eqref{metric} and \eqref{Demianski} are locally isometric can now be seen as follows: we start with the Demianski metric \eqref{Demianski} and rescale it using the scaling transformation \eqref{scaling} with $\a^2=1-\ka_D=1-\lla b_D$.  The rescaling of the Demianski geometry with angular momentum $a_D$ leads to the metric \eqref{metric} with $\ep=1$  and
\begin{equation}
 a=\frac{a_D}{1-\lla a_D^2}.
\end{equation}
 Therefore the $m=0$ Demianski's geometry is locally isometric to the geometry of \eqref{metric}. The question to whether the  $m=0$ Demianski geometry is isomorphic to our rotating solution is more complicated because the value of $b$ affects the ranges other coordinates, in particular of $p$ and $\vp$. For both metrics $\vp=0$  is identified with $\vp=2\pi$ and in addition the range of $p$ is restricted by $|p|\leqslant a$. Thus whereas the metric \eqref{metric} is defined for standard values of $\th$ and $\vp$, for the embedded Demianski geometry $\vp=0$ is idenfified with $\vp=2\pi\sqrt{1-\ka_D}$ and $p$  is restricted by $|p|\leqslant a \sqrt{1-\ka_D}$.

\begin{center}
  \includegraphics[scale=0.55]{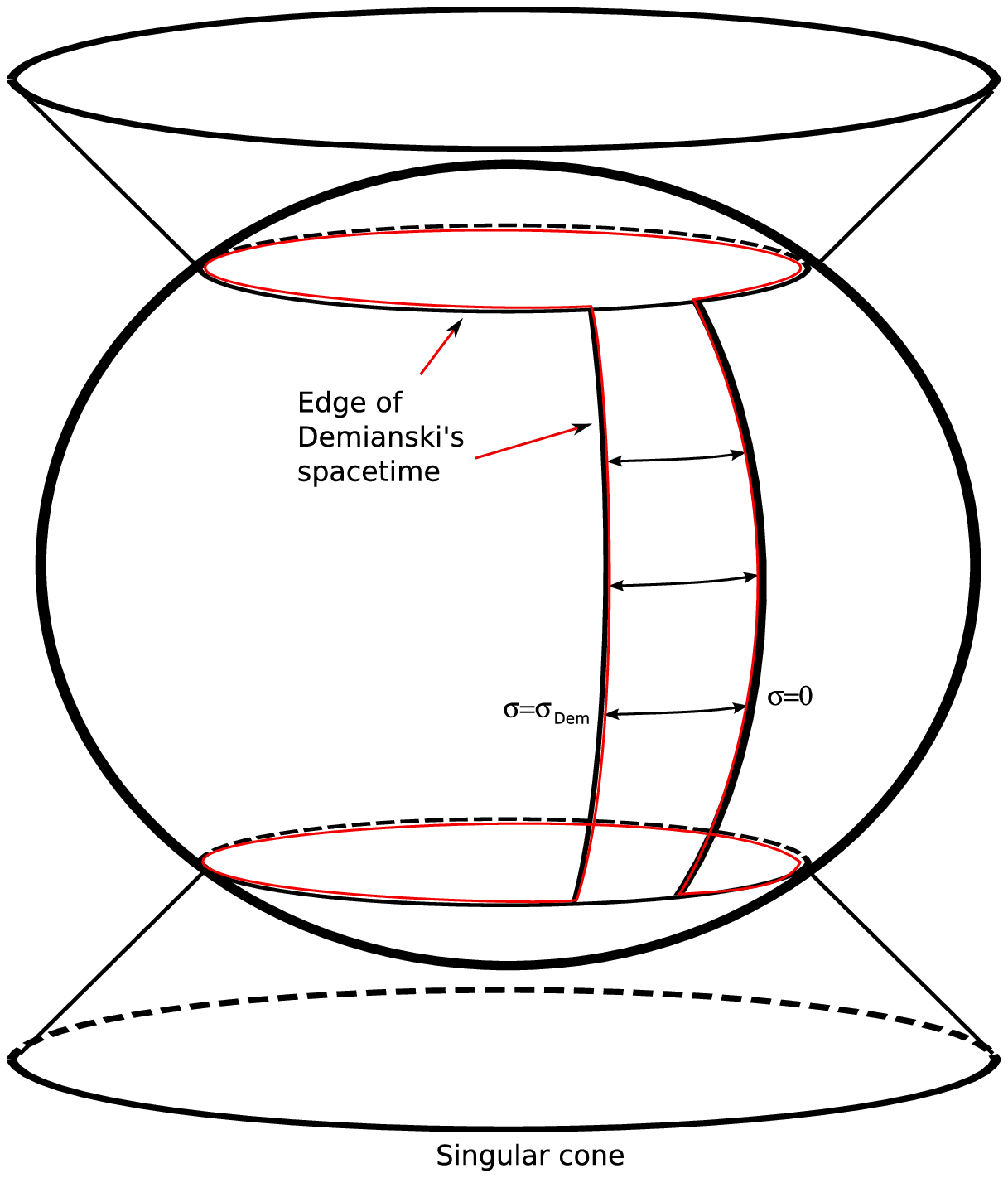}
\end{center}
{\it Fig. 2. Embedding of the Demianski space-time in the space-time \eqref{metric}.\\[7pt]}

Due to $\P=\tfrac{1}{\a^4}\, \P_D$, the latitudes covered by the $m=0$ Demianski are just those outside the conical horizon surface Eq. \eqref{conical_horizon} (see Fig.2). Moreover, the Demianski's horizon, i.e. the surface $(r_{h})_D=1/{\sqrt{\lla}}$ in Demianski's coordinates,  corresponds to $r_{h}=(1/{\sqrt{{\lla}}})\cdot (1-\ka_D)^{-1/2}$ in our coordinates, and is always located further than the spherical horizon of our metric,  $r_H=(1/{\sqrt{\lla}})\cdot (\tfrac{1}{2}+\tfrac{1}{2}\sqrt{1+\ka}\, )^{1/2}$  (the spherical horizon, \eqref{spherical_horizon}).

We finally note, that other rotating solutions of the Einstein equations that might serve as models for the interior of rotating compact objects, but having stress-energy tensors different from a simple $\La$ term, have been published \cite{Burinski,Dymnikova}. Burinski et al. \cite{Burinski} have investigated a class of spherically symmetric Kerr-Schild type space-times, which can be transformed to rotating solutions by the Newman-Janis ``complex trick'' \cite{Stephani}.
While the transformed Schwarzschild solution is just the Kerr solution, the transformed de Sitter solution is a rotating space-time that is not a solution of vacuum field equations with a $\La$ term. This solution and the similar solution of Dymnikova \cite{Dymnikova} are interesting because of their continuity on the junction surface between the interior and the exterior Kerr regions. Our solution is distinguished by having only a vacuum energy and the appearance of a vortex-like core.

\section{Behavior near to the axis of rotation}
$\qquad$ As noted above the metric \eqref{metric} is plagued by time-like closed curves for near to the conical horizon. Closed time-like curves are extremely pathological from the point of view of quantum mechanics. Inside the region with CTCs there is the conical singular surface, the interior of which cannot be interpreted as a space-time at all (signature is $+++-$).  As in the condensate vortex picture for space-time spinning strings \cite{CHM04} where the condensate density goes to zero very near to the axis of rotation (where CTCs reside), it may be a reasonable guess that physical space-time undergoes a phase transition where closed time-like curves appear in our interior solution for a rotating compact object. Thus we expect the region close to the axis of rotation of the compact object not to be described by the solution Eq. \eqref{metric} anymore. The transition to a new phase should take place at least at the boundary of the region with CTCs. Following the ansatz introduced in ref. \cite{CHHLS} that the two different space-times on either side of this phase boundary represent classical solutions of the Einstein eq's. whose metrics match at the boundary, it is reasonable to guess that the G\"odel-like (e.g. the Som-Raychaudhuri) space-time discussed in Ref. \cite{CHM04}) may be a good approximation for the space-time metric in the part of the space-time nearest the axis of rotation of the compact object, at least near z equal 0 in Fig 1. This metric would be applicable inside the critical radius where the local speed of frame rotation is equal to the speed of light. Since outside this critical radius the metric \eqref{metric} is locally isometric to ordinary de Sitter space, we can also conclude that the angular momentum of the compact object will actually be carried by vortices concentrated near to the axis of rotation. As a result of this concentration of vortices a solid body-like rotation of the space-time near to the axis appears. It is amusing that a giant vortex near the axis of rotation also appears in the theory of rapidly rotating superfluid droplets \cite{Zeldovich}.

The solutions of the equations of motion for particles in a G\"odel-like space-time are well known (see e.g. \cite{NovelloSoaresTiomno}). In general this flow of particles (including photons) will be collimated along the distinguished direction of the space-time. The radius of the G\"odel-like phase, defined though the circumference of circles with center on the axis of rotation of the compact object, should not exceed the radius where the frame rotation velocity equals the speed of light. In our situation this is where closed time-like curves first appear in our solution as the axis of rotation is approaced; i.e. the inequality in eq. \eqref{CTCs} becomes an equality. The G\"odel-like region acts as a cylindrical trap for free particles: every particle (null- or time-like geodesic) emitted close to the axis of rotation will be confined to move inside the critical radius.  On the other hand particles are free to move parallel to the axis, so that for slow rotation of our compact object the flow of particles along the axis of rotation will be highly collimated.

\section{Conclusion}
$\qquad$ In summary, the metric \eqref{metric} suggests a model for the interior space-time of a rotating compact objects that is de Sitter-like except for regions near to the axis of rotation. Although our model does not provide an exact solution, we believe that our model correctly identifies some remarkable qualitative features of physical compact objects. In particular, the existence of a conical horizon forces the de Sitter region to have an effectively toroidal topology and implies that another phase of space-time- perhaps G\"odel-like - will fill the hole of the torus. This may be a crucial clue as to the origin of highly collimated jets from compact astrophysical objects.

It is interesting that in contrast with the non-rotating case where general relativity only fails in a thin layer near to the spherical event horizon, we find that the external vacuum (Kerr) space-time will need to be drastically modified for high latitudes, perhaps even to large distances from the compact object. These corrections to the Kerr solution could produce observable effects for stars orbiting the compact object in polar-like orbits.

Remarkably we find that general relativity fails in a more global way in \emph{the case of rotating objects}. This global failure of general relativity is fundamentally due to the incompatibility of general relativity and quantum mechanics.  Finally we should note that our metric \eqref{metric} might also serve as a model for the large scale structure of our universe inside the de Sitter horizon, where there are hints from observations of the large scale anisotropy of the cosmic microwave background that the universe might be rotating \cite{Chapline}.\\[10pt]

{\noindent One of the authors (GC) is very grateful for numerous discussions with
Pawel Mazur.}



\end{document}